\documentclass[%
floatfix, 
 reprint,
 amsmath,amssymb,
 aps,
]{revtex4-2}
\usepackage[top=2cm,bottom=2cm,left=1.5cm,right=1.5cm,columnsep=0.7cm]{geometry}
\usepackage{xparse,xcoffins}

\mathchardef\mhyphen="2D

\ExplSyntaxOn
\NewCoffin\imagecoffin
\NewCoffin\labelcoffin

\keys_define:nn { alex/label }
 {
  label   .tl_set:N = \l_alex_label_tl,
  labelbox .bool_set:N = \l_alex_label_box_bool,
  labelbox .default:n = true,
  fontsize .tl_set:N = \l_alex_label_size_tl,
  fontsize .initial:n = \footnotesize,
  pos .choice:,
  pos/nw .code:n = \tl_set:Nn \l_alex_label_pos_tl { left,up },
  pos/ne .code:n = \tl_set:Nn \l_alex_label_pos_tl { right,up },
  pos/sw .code:n = \tl_set:Nn \l_alex_label_pos_tl { left,down },
  pos/se .code:n = \tl_set:Nn \l_alex_label_pos_tl { right,down },
  pos/n .code:n = \tl_set:Nn \l_alex_label_pos_tl { hc,up },
  pos/w .code:n = \tl_set:Nn \l_alex_label_pos_tl { left,vc },
  pos/s .code:n = \tl_set:Nn \l_alex_label_pos_tl { hc,down },
  pos/e .code:n = \tl_set:Nn \l_alex_label_pos_tl { right,vc },
  pos .initial:n = nw,
  unknown .code:n   = \clist_put_right:Nx \l_alex_label_clist
                       { \l_keys_key_tl = \exp_not:n { #1 } }
 }
\clist_new:N \l_alex_label_clist
\box_new:N \l_alex_label_box
\box_new:N \l_alex_label_image_box

\NewDocumentCommand{\xincludegraphics}{O{}m}
 {
  \group_begin:
  \tl_clear:N \l_alex_label_tl
  \clist_clear:N \l_alex_label_clist
  \keys_set:nn { alex/label } { #1 }
  \tl_if_empty:NTF \l_alex_label_tl
   {
    \alex_includegraphics:Vn \l_alex_label_clist { #2 }
   }
   {
    \SetHorizontalCoffin\imagecoffin
     {
      \alex_includegraphics:Vn \l_alex_label_clist { #2 }
     }
    \SetHorizontalCoffin\labelcoffin
     {
      \raisebox{\depth}
       {
        \bool_if:NTF \l_alex_label_box_bool
         { \fcolorbox{white}{white}{\l_alex_label_size_tl\l_alex_label_tl} }
         { \l_alex_label_size_tl\l_alex_label_tl }
       }
     }
    \SetVerticalPole\imagecoffin{left}{35pt+\CoffinWidth\labelcoffin/2}
    \SetVerticalPole\imagecoffin{right}{\Width-3pt-\CoffinWidth\labelcoffin/2}
    \SetHorizontalPole\imagecoffin{up}{\Height--11.5pt-\CoffinHeight\labelcoffin/2}
    \SetHorizontalPole\imagecoffin{down}{3pt+\CoffinHeight\labelcoffin/2}
    \use:x{\JoinCoffins\imagecoffin[\l_alex_label_pos_tl]\labelcoffin[vc,hc]} 
    \TypesetCoffin\imagecoffin
   }
   \group_end:
 }
\NewDocumentCommand{\setlabel}{m}
 {
  \keys_set:nn { alex/label } { #1 }
 }

\cs_new_protected:Nn \alex_includegraphics:nn
 {
  \includegraphics[#1]{#2}
 }
\cs_generate_variant:Nn \alex_includegraphics:nn { V }

\ExplSyntaxOff

\usepackage{bbm} 

\usepackage[T1]{fontenc}
\usepackage[utf8]{inputenc}
\usepackage[english]{babel}

\usepackage{amsfonts,amssymb,amsmath,array}
\usepackage{bm}
\usepackage{dcolumn}
\usepackage[output-decimal-marker={.}]{siunitx} 

\usepackage[caption=false]{subfig}
\usepackage{comment}

\usepackage{graphicx} 
\usepackage{listings} 
\usepackage{ulem}
\usepackage{microtype}
\usepackage{color}
\usepackage{float}

\usepackage{hyperref}
\hypersetup{
    colorlinks=true,
    linkcolor={blue}, 
    filecolor=magenta,   
    citecolor={blue}, 
    urlcolor=blue, 
}

\newcommand{\br}{{\bf r}}

\newcommand{\bF}{{\bf F}}
\newcommand{\bn}{{\bf n}}

\usepackage{enumitem}

\begin{document}

\preprint{APS/123-QED}

\title{
Extinction and coexistence 
in a
binary mixture of proliferating 
motile disks
}

\author{Alejandro Almod\'ovar}
\email{almodovar@ifisc.uib-csic.es}
\author{Tobias Galla}
\email{tobias.galla@ifisc.uib-csic.es}
\author{Crist\'obal L\'opez}
\email{clopez@ifisc.uib-csic.es}
\affiliation{IFISC, Instituto de F\'isica Interdisciplinar
y Sistemas Complejos (CSIC-UIB), Campus Universitat de les
Illes Balears, E-07122 Palma de Mallorca, Spain}

\begin{abstract}
A binary mixture of two-different-sizes
proliferating motile disks is studied.
As growth is space-limited, we focus 
on the conditions such
that there is coexistence 
of both large and small disks,
or dominance of the larger disks.  
The study involves 
systematically varying 
some system parameters, such as
diffusivities, growth rates, and 
self-propulsion velocities.
In particular, we
demonstrate that 
diffusing
faster
confers a competitive advantage, 
so that larger 
disks can in the long
time  coexist or even
dominate to the smaller ones.
In the case of  self-propelled disks
 a coexistence regime
is induced by the activity
where the two types of 
disks show the same spatial
distribution: both phase 
separated or both homogenously
distributed in the whole system.

\end{abstract}

\maketitle

\section{Introduction}
\label{Sec:intro}

Motivated by the collective dynamics in biological
phenomena such as wound healing \cite{XTrepat2020,XTrepat2021,Sepulveda}, 
tissue formation \cite{Angelini2011}, the expansion
of tumours \cite{Zapperi}, or the dynamics of bacterial populations \cite{Szabo2006},
there is  growing interest in the study of proliferating
motile matter \cite{Hallatschek2023,Tang2023}. 
These  are often modelled as interacting
particle systems \cite{Kim2022,Fu2019,Sepulveda,Zapperi}, 
and the number of particles may not be constant
in time due to
processes such as birth and death. 
Coupled with individual movement one can expect
new  emergent 
properties. For example, we
recently introduced a simple model of 
{\it proliferating} motile finite-size
particles. Specifically, in \cite{Almodovar2022} 
we studied both systems of 
passive disks
(where motility has its origins in a thermal bath)
 and 
self-propelled 
disks \cite{Romanczuk2012,tenHagen2011,Fily2012,digregorio2018,caprini2020} 
subject to reproduction and death. 
We
analysed different emerging structural phases 
[liquid, hexatic, solid and motility-induced
phase separation (MIPS)]
 \cite{Bernard2011,Kapfer,Farid1981,Nelson1979,
Cates2015,Buttinoni2013,caprini2020}. 
These phases result as 
a consequence of the disks
filling the available space, 
and the phase  that is realised 
depends on parameters such as 
the birth and death rates, 
and the motility.

In \cite{Almodovar2022} we focussed on 
a two-dimensional collection
of disks (we will refer to these also as particles),
which were all taken 
to be identical, in particular,
they have the same size. This is 
not realistic in many biological applications. 
In the current work we analyse
the influence on some properties of the system  when
relaxing this condition, and
consider a binary mixture of 
disks of two different sizes \cite{Biben1991,Frenkel1992}.
In the standard equilibrium system
of hard disks (without birth-death events
or activity)
the two-sizes binary mixture is
known to have important effects
leading for example to the dissapearence
of the hexatic phase even 
for very low concentration
of small disks \cite{Russo2017}.
Thus we expect new 
behavior in a
non-equilibrium model of a binary mixture
with birth and
death dynamics.


We start with a similar concentration 
for both types (or species) of disks, 
initially randomly distributed
in a two-dimensional space. As the system evolves,
 particles move, die and reproduce,
  eventually filling up space. 
 Due to the death processes it 
 is possible for one species to go extinct.
In reproduction events a new 
particle is created in the system,
with the same size as the parent,
and placed close to it if there 
is enough room.  

As in many real-world systems,
reproduction in our model
is limited by the available space 
\cite{larwood1979,Jackson1977,Ponczek2001}. 
Under identical conditions,  
the smaller disks have a 
higher a-priori chance to reproduce, and 
therefore to persist in the long-run. 
Even though the availability of 
space puts the larger particles at
a disadvantage \cite{SuchismitaDas2020,Reversat2020}, 
we find that,
depending on motility and demography, 
coexistence of both species
can occur.
In some other cases
we find that the larger particles
can dominate the system,
i.e. their number is much larger than that of the smaller 
particles, typically more 
than $90\%$ of the total particle number.

The main questions 
we address are  what type of 
disk dominates the system
in the long-run, under what conditions both types co-exist, 
and, in the important case of
activity-induced phase separation,
what the resulting spatial structure is.
We are interested 
in the influence of the birth and death rates 
and the motilities of the particles 
on the outcome. 
In particular, we ask 
which population survives
when the two species compete
(due to reproduction limited by
size)
with different diffusivities
\cite{Pigolotti2014,Heinsalu2012},
and we show results that support
the prevalence of the fast diffusing type of particle.
Our overall aim is to 
characterise the system behaviour 
for different choices of the model parameters.

In the case of active 
motion \cite{Marchetti2013,Bechinger2016,Hakim2017} 
we also study what the
conditions are under which
the spatial distribution of the particles
presents coexisting dilute and dense phases, i.e.,
the MIPS regime. This is for 
example motivated by studies 
hypothesing that 
phase heterogeneity 
due to MIPS
can trigger a transition from swarming 
behaviour to biofilm formation 
in some type of bacteria  \cite{Grobas2021}. 
We analyse the role of
this heterogeneity for the dominance
dynamics of the binary mixture.

Throughout our analysis, we will mostly concentrate 
on systems in which the diameter of one type of disk is $20 \%$ larger than the other, but we will also discuss other  size ratios 
\cite{Bommineni2019,SampedroRuiz2019,SampedroRuiz2020,Kumar2021}.

The remainder of the paper 
is organised as follows.
In Sec.~\ref{Sec:model}, we present
the model of mobile disks
undergoing birth and death dynamics.
The main outcomes  are
presented in Sec.~\ref{Sec:results}.
A summary and discussion of the findings 
is  contained in Sec.~\ref{Sec:summary}.


\section{Model and numerical algorithm}
\label{Sec:model}

The model is similar to that in \cite{Almodovar2022}
but with particles of two different sizes. 
We consider a two-dimensional system of $N(t)=N_L (t) + N_S (t)$
interacting disks with diameters $\sigma_S$ and $\sigma_L$
such that $\sigma_S < \sigma_L$
 ($S$ and $L$ stands for `small' and `large', respectively). The 
 particle numbers can change in time, 
 due to birth and death events, as explained below.
We consider the overdamped limit and take the friction
 coefficient to be equal to unity 
 for both species. 
 The motion of disks is then as follows,
\begin{equation}
\dot{\br_i}= {\bF}_i + {\bF}^{act}_i + \sqrt{2D_i} \boldsymbol{\zeta}_i (t), \ \ i=1,...,N(t).
\end{equation}
If disk $i$ is of the small type
then $D_i = D_S$, and if it is of 
the large type then $D_i = D_L$. The variables
$\{\boldsymbol{\zeta}_i\}$ are independent Gaussian noise
vectors satisfying 
$\langle \boldsymbol{\zeta}_i \rangle= 0$, 
$\langle {\zeta}_{i,a} (t) { \zeta}_{j,b} (t') \rangle = \delta_{i j} \delta_{a b}\delta (t-t')$ 
($a$ and $b$ are the entries  of the two-component vectors
 $\boldsymbol{\zeta}_i$ and $\boldsymbol{\zeta}_j$). 
No Einstein relation is assumed,
and $D_L$ and $D_S$ are taken
as parameters of the model.
The finite size of the disks is simulated using
a truncated Lennard-Jones potential  
so that the force on particle $i$ 
resulting from the 
interaction with 
the rest of particles
is $\bF_{i} = -{\bf \nabla}_i \sum_{i\neq j}  U(|\br_i -\br_j|)$,
where the potential is given by (with 
$r=|\br_i -\br_j|$)
\begin{equation}
U(r) = 4 \varepsilon [(\frac{\sigma_{ij}}{r})^{12}
- (\frac{\sigma_{ij}}{r})^6] + \varepsilon,
\end{equation} 
if $r< 2^{1/6} \sigma_{ij}$, 
and $U(r)=0$ if $r > 2^{1/6} \sigma_{ij}$.
The quantity $\sigma_{ij}>0$ is defined 
from the Lorentz-Berthelot 
rule as $\sigma_{ij}=(\sigma_i + \sigma_j)/2$, 
where $\sigma_i$ and $\sigma_j$ are 
the diameters of disks $i$ and $j$, respectively. Thus,
$\sigma_{ij}$ reflects the effective distance between the two disks. The parameter
$\varepsilon$ is an energy scale \cite{LB_rule_Book1,LB_rule_Book2}.

In some of our  numerical experiments particles are self-propelled. 
We model this using active forces 
\begin{equation}
\bF^{act}_i = v_i \bn[\theta_i (t)]
\label{eq:forceact}
\end{equation} 
of constant modulus $v_i = \{v_L, v_s \}$ 
(typically called activity or velocity)
 and
with a direction given by the unit vector
$\bn (\theta_i) = ( \cos \theta_i, \sin \theta_i )$.
The angle $\theta_i$ for disk $i$ performs diffusive motion,
$\dot{\theta}_i (t) = \sqrt{2 D_r} \eta_i (t)$.
The term $\eta_i$ represents a zero-mean Gaussian noise with
$\langle \eta_i (t) \eta_j (t')\rangle = \delta_{ij} \delta (t - t')$. 

In addition to movement and interaction,
disks may randomly self-replicate or die, so that
the number of disks of each type, $N_L (t)$ and $N_S (t)$,
can change with time. These events occur as 
follows (see \cite{Almodovar2022} for
further details of the  algorithm):
\begin{enumerate}[topsep=5pt,itemsep=-1ex,partopsep=5ex,parsep=5ex]
\item Death occurs as a Poisson process. Each existing particle 
dies with per capita rate $\delta$.
 Particles that die are  removed from the system.
\item  Potential births  are triggered 
with per capita rate $\beta_L$ for large disks,
and $\beta_S$ for small disks. The diameter of the potential offspring 
is identical to that of the parent. The birth event only occurs if there
is sufficient space around the parent particle to place the
offspring without overlapping
with any other disk. If there is no space, no birth event occurs. 
This means that not all potential reproduction events complete.
\end{enumerate}
We always assume that the growth rates $\beta_S$ and $\beta_L$ are 
larger than the death rate (the latter is taken 
to be equal for all the disks), i.e. $\beta_L,\beta_S >\delta$.
At long times the system reaches a stationary state. 
The numbers of disks of each type in this state are 
such that the mean effective birth rate for each 
species is equal to the death rate. 

In the next sections
we study this steady state while varying one 
or two model parameters at a time
(e.g., the diffusivity of both 
types of particles, the growth rates or activity).
Our main objective is to 
study what type of disk
dominates, and under what conditions there
is coexistence. 


\section{Results}
\label{Sec:results}

We consider a two-dimensional
box of length $L_s=150$ with
periodic boundary conditions,    
$\delta=0.01$,
and for the most part
$\sigma_S=1.0$ and $\sigma_L=1.2$,
supplemented by some discussion
of other choices of the diameters.
For simplicity, we set
$\varepsilon=1.0$ and $D_r = 1.0$. Simulation 
results are independent of $\delta$
whose role is mostly to set the time
scale needed to reach the steady state 
\cite{Almodovar2022}.
We start with $250$ particles of each size,
together occupying around $8\%$ of the total area.
We  compute the packing fraction 
for particles of type. We write these
as  
$\phi_{\alpha}(t)=N_{\alpha}(t)\pi (\sigma_{\alpha}/2)^2/L_s^2$
for $\alpha\in\{L,S\}$,
where $N_{\alpha}(t)$ is the number of 
particles of type $\alpha$ at time $t$.


We study passive 
particles in Sec.~\ref{sub:pp}. 
Systems of active particles 
are discussed in Sec.~\ref{sub:active}.

\subsection{Passive particles}
\label{sub:pp}

We set  $\mathbf{F}_i^{act}=0$,
and study the effects of the parameters of the
demographic dynamics, and 
of the diffusivities separately.

\subsubsection{Effects of the birth and death rates}
\label{sub:demo}

\begin{figure*}[hbt!]
  \centering
    \includegraphics[width=.95\linewidth]{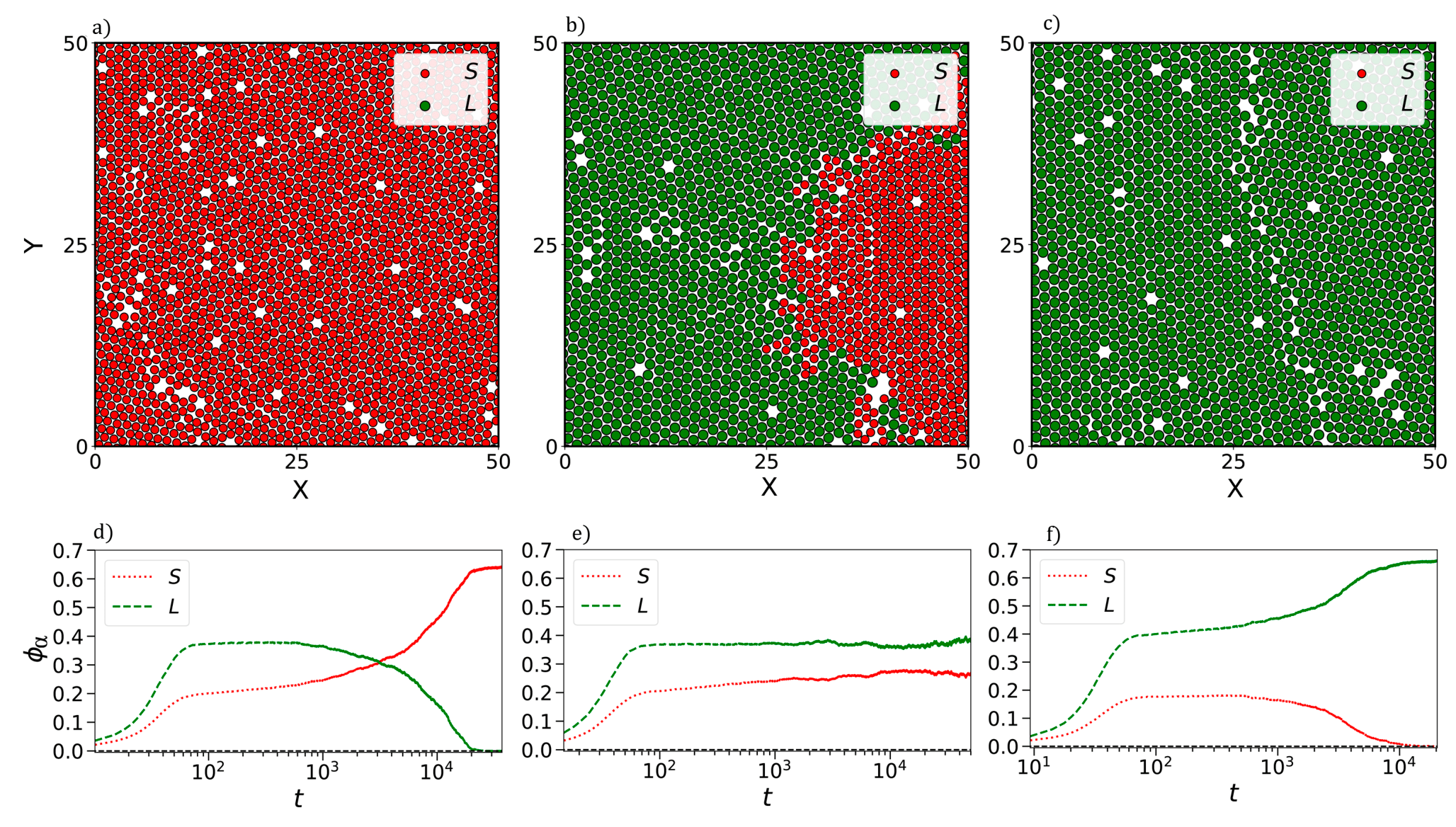}
\caption{Panels in the first row (a-c) show snapshots 
  of the spatial distribution of passive disks in space at long times
  (only a part of the full size $L_s=150$ is shown). 
  Panels in the second row (d-f) correspond 
  to the temporal evolution of the packing fraction
  starting with a random configuration of $250$ disks of each type. 
  In each column, we have used different values of $\beta_L$, at fixed $\beta_S=0.1$, and $\beta_L = 0.114, 0.118, 0.13$ from left to right. The remaining  
parameters are $\delta=0.01$, $D=0.001$, $\sigma_L/\sigma_S=1.2$.}
\label{Fig:Plots_Birth}
\end{figure*}

We fix the diffusivities to 
low values (i.e. the time scale
of a disk to move a distance 
of the order of several diameters is  
large compared to its
typical lifetime) $D_L=D_S=0.001$,
to focus on   
the effects of the birth and death 
dynamics.
Since reproduction 
is limited by the available space, we naively
expect only the smaller disks 
to be present in the long-run when the
raw birth rates of both types 
are similar.
However, the outcome may 
be different if the birth 
rate for the larger particles is much
larger than that of the 
smaller type. This is indeed what 
we observe in the numerical 
experiments shown in Fig.~\ref{Fig:Plots_Birth}.

In the upper row of Fig.~\ref{Fig:Plots_Birth} we 
plot snapshots of the spatial distribution
of disks in the stationary state. 
Smaller particles shown
in red,  larger  ones in green. 
The growth rate of the 
small particles is lower than 
that of the larger particles in all three panels,
but the ratio $\beta_L/\beta_S$ 
increases from left to right. In 
panel a) we observe 
the extinction of the larger particles,
in 
panel b) we find balanced
co-existence throughout the simulation,
and in panel c) the
ratio $\beta_L/\beta_S$ is sufficiently 
large for the larger  disks 
to fully dominate (the smaller particles go extinct). 
In the lower panels 
of Fig.~\ref{Fig:Plots_Birth} 
we plot the corresponding 
time evolution of the packing 
fractions (the snapshots in 
the upper panels are taken at 
the final time of the lower ones).
These time series confirm the extinction 
of the larger type of particle
[panel d)], 
co-existence [panel e)] and the extinction
of the smaller type [panel f)], respectively. 
In all
three situations, and regardless
of the final fate of the larger particles,
 there 
is initially a
faster increase of the 
packing of the larger particles. 
This is because
$\beta_L > \beta_S$ in all 
panels in Fig.~\ref{Fig:Plots_Birth}

\begin{figure}
\includegraphics[width=0.99\linewidth]{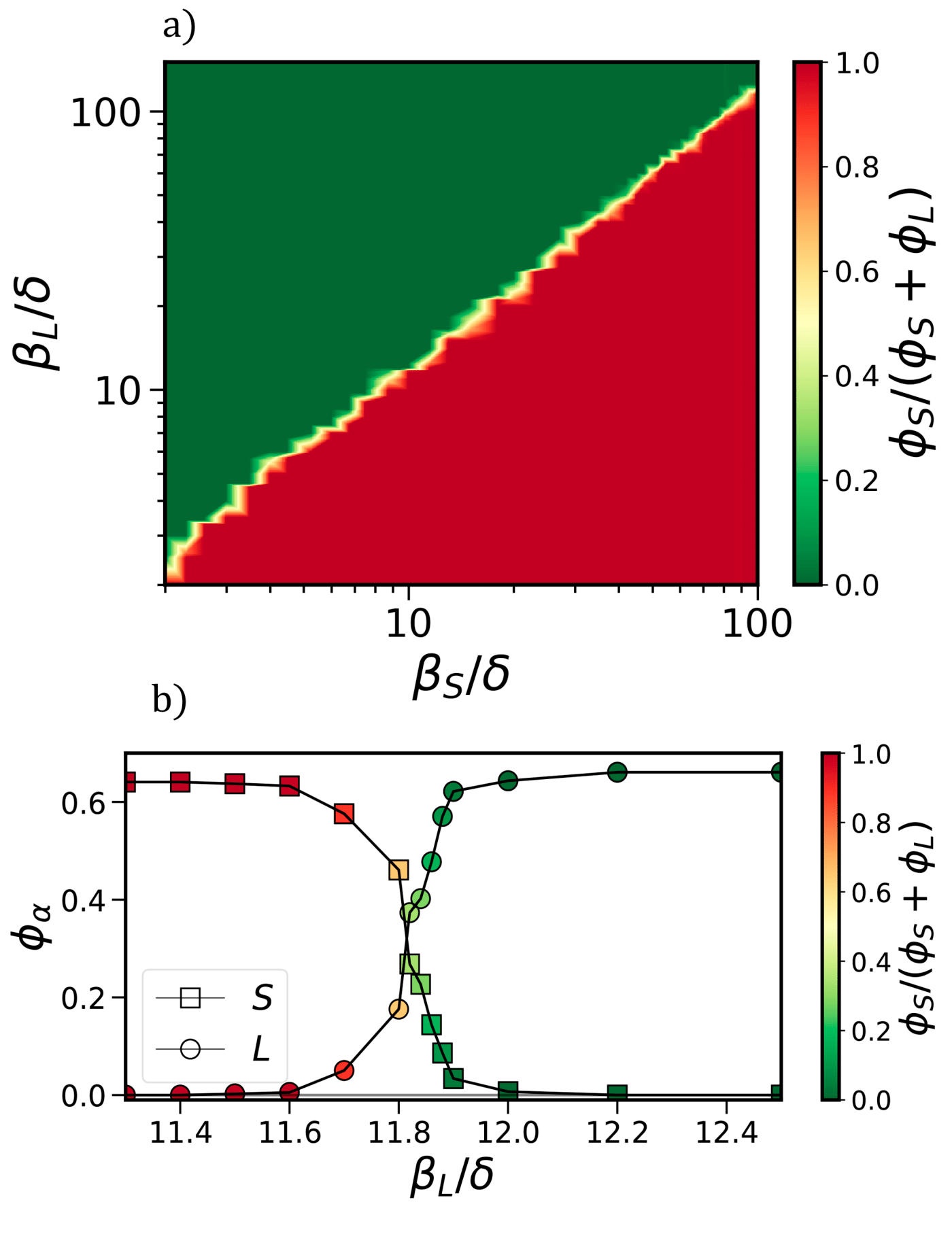}
\hspace*{-0.3cm}

\caption{Coexistence diagram for the system of passive particles 
in the plane spanned by the birth rates $\beta_S$ and $\beta_L$. 
Panel a): color indicates where in parameter space either 
type of particle dominates, or if there is co-existence. 
Panel b): Vertical cut in the phase diagram in (a) 
for $\beta_S/\delta = 10 $ showing
the packing fractions for 
both types of disks. The colors of
the symbols represent the fraction of 
small particles in the system. $\delta=0.01$, $\sigma_S=1.0$, and $\sigma_L=1.2$.}
\label{Fig:Diagram_Birth}
\end{figure}

In Fig.~\ref{Fig:Diagram_Birth}a) we plot
the phase diagram in 
the plane of birth rates
 for the two species
(each divided by the death rate
$\delta=0.01$). The colored heatmap indicates the
normalised packing fraction $\phi_S/(\phi_S+\phi_L$) 
of the smaller disks. 
In Fig.~ \ref{Fig:Diagram_Birth}b) we show 
the respective packing fractions of both types
when varying $\beta_L$,
for  fixed  $\beta_S$. This corresponds 
 to a vertical 
cut of the phase diagram in panel a). 
The dominance of the smaller 
type at low  $\beta_L$,
co-existence at intermediate values, and
dominance of the larger 
 disks at high $\beta_L$ are clearly visible.


Coexistence of both species occurs in a very 
limited region of the diagram,
around a small area near a specific line 
in the $(\beta_S,\beta_L)$ plane.
An approximate characterisation of this 
coexistence curve can be obtained from a
simple description of the  
dynamics in terms of 
rate equations.
To construct these equations
with proper parameters, we first consider 
only one type of 
disk in isolation, neglect 
fluctuations and any notion of space.
 We 
use a   Lotka--Volterra model
 to describe the combination of death events
  and growth limited by volume exclusion \cite{Khalil2017}, 
\begin{equation}
 \frac{dN}{dt} = - \delta N + {\hat \beta} N \left(1 - \frac{N}{N_{\rm max}}\right),
 \label{eq:growth}
\end{equation}
where the term in the brackets ensures 
that the growth dynamics stops 
as $N$ reaches $N_{\rm max}$;
$\delta$ is the death rate, and  
$\hat \beta$ is an effective growth rate.

This effective growth rate is a fitting 
parameter introduced in 
this very simple description 
to take into account that the actual  birth rate 
is limited by the available space,
and therefore not equal to the raw birth rate $\beta$. 
We take $\hat \beta = a \beta$,
where $a$ is
an unknown adimensional coefficient.
In \cite{Almodovar2022} we observed that the
steady number of disks 
in a single-species model
depends in a
non-trivial way on the birth and
death rates, the size of the particles,
and on the diffusion coefficient. 
Thus, in this
description
we  use $a>0$ and $N_{\rm max}$ 
as  fitting parameters.

We rewrite Eq. (\ref{eq:growth}) in the standard form
\begin{equation}
dN/dt= r N (1-N/K),
\label{eq:growth2}
\end{equation}
with
$r=a \beta - \delta$.
The carrying capacity 
$K = N_{\rm max} (1 - \frac {\delta}{a \beta})$
is the effective long-time
number of disks in the system.

\begin{figure}
\includegraphics[width=0.95\linewidth]{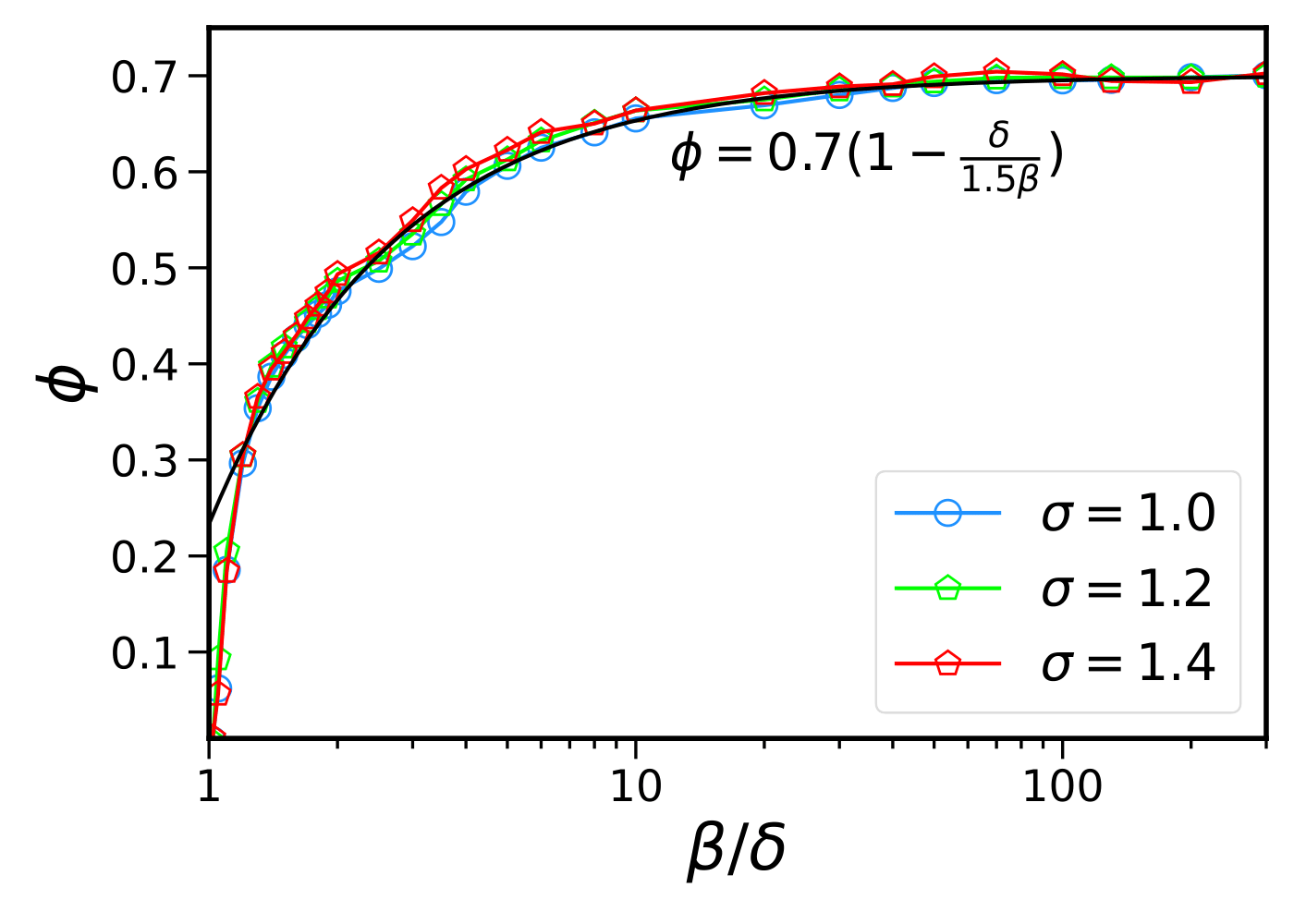}
\caption{Long-time average packing fraction,
$\phi_{\infty}$,
 for a system
of only one type of disks as a function 
of the rate $\beta$ and $\delta=0.01$.
The plots correspond to different disk
sizes $\sigma$ as the symbols indicate.
The black line corresponds to the best fitted curve
for $a=1.5$, and $\phi_{\rm max}=0.7$.}
\label{Fig:Lotka}
\end{figure}

To verify the validity of the  
logistic-growth approach we 
have carried out simulations of the model 
in which all particles have 
the same size,  using a small
diffusion coefficient, $D=0.001$.
In the long-time we compute the packing fraction
$\phi_{\infty} = N(t\to \infty) \pi (\sigma/2)^2 / L_s^2$.
We do this for several different disk sizes
($\sigma=1.0, 1.2, 1.4$)
and for some values of $\beta$ to check
the relationship $\phi_{\rm stat} = K \pi (\sigma/2)^2 / L_s^2 = 
N_{\rm max} \pi (\sigma/2)^2 / L_s^2 (1 - \frac {\delta}{a \beta})$
(the stationary packing fraction of the Lotka\textendash Volterra equation).
The results are shown in Fig.~\ref{Fig:Lotka}. 
There is a good collapse of
all the plots for the different values of $\sigma$,
and the best fit to the
expression of $\phi_{\infty}$ is obtained
for $\phi_{\rm max}= N_{\rm max} \pi (\sigma/2)^2 / L_s^2 =0.7$, $a=1.5$.
This confirms that the 
logistic description
with suitable effective parameters
can be used to describe the stationary
state with 
a single type of
particle.

The Lotka--Volterra model can be 
generalised to the case of a binary
mixture  of disks. This
approximate description
can be useful to study 
the influence
of one type of disks on the other,
in particular, attending to their
different sizes and growth rates. 
We assume the following Lotka\textendash Volterra 
competition dynamics
between the two species 
(which we label again $S$ and $L$ 
for `small' and `large' disks,
 respectively) \cite{Edelsteinbook}:
\begin{eqnarray}
\frac{dN_L}{dt}= - \delta N_L +  a\beta_L \left(1 - \frac{N_L + \alpha_L N_S}{N_{L,{\rm max}}}\right) N_L, \nonumber \\
\frac{dN_S}{dt}= - \delta N_S +  a\beta_S \left(1 - \frac{N_S + \alpha_S N_L}{N_{S,{\rm max}}}\right) N_S.
\label{eq:competition}
  \end{eqnarray}
The term $\alpha_L N_S$ ($\alpha_S N_L$) is
the decline of the growth rate of $N_L$ ($N_S$)
due to the presence of the smaller (larger) disks. 
The quantities 
$N_{S,{\rm max}}$ and $N_{L,{\rm max}}$ are the
number of particles for either species 
at which no further growth can occur.

The dependence of the interaction terms on the ratio
of disk sizes becomes more apparent if we write this
equation in 
terms of packing fractions:
\begin{eqnarray}
\frac{d\phi_L}{dt}= - \delta \phi_L +  a\beta_L \left(1 - \frac{\phi_L + \alpha_L \frac{\sigma_L^2}{\sigma_S^2}\phi_S}{\phi_{L,{\rm max}}}\right) \phi_L, \nonumber \\
\frac{d\phi_S}{dt}= - \delta \phi_S +  a\beta_S \left(1 - \frac{\phi_S + \alpha_S \frac{\sigma_S^2}{\sigma_L^2}\phi_L}{\phi_{S,{\rm max}}}\right) \phi_S.
\label{eq:competitionpacking}
  \end{eqnarray}

Now $\phi_{\alpha,{\rm max}} = N_{\alpha,{\rm max}}\pi (\sigma_\alpha/2)^2/L_s^2$ ($\alpha=L,S$).
We assume that $\alpha_L$ and $\alpha_S$
depend on the demographic parameters of the system
$(\beta_L, \beta_S)$, and as detailed in
 Appendix \ref{ApA}. Fitting to simulation data we find 
\begin{eqnarray}
\label{eqn:alfas}
    \alpha_L &\approx& \left(\frac{\beta_S}{\beta_L}\right)^{1/4}, \nonumber \\
 \alpha_S &\approx& \left(\frac{\beta_L}{\beta_S}\right)^{1/4}.
\end{eqnarray}

These relations characterise 
the tradeoff between geometry and
demography (reproduction)
in the competition dynamics. 
For example the coefficient 
in the growth rate for the larger disks
$\alpha_L \frac{\sigma_L^2}{\sigma_S^2}=
\frac{\sigma_L^2}{\sigma_S^2}
\left(\frac{\beta_S}{\beta_L}\right)^{1/4}$
increases  proportionally to
$(\sigma_L/\sigma_S)^2$,
while its 
dependence on the relative
birth rates is much weaker and
only scales as 
$(\beta_S/\beta_L)^{1/4}$.


\subsubsection{The role of diffusion}
\label{sub:diff}

\begin{figure*}[hbt!]
  \centering
  \includegraphics[width=.99\linewidth]{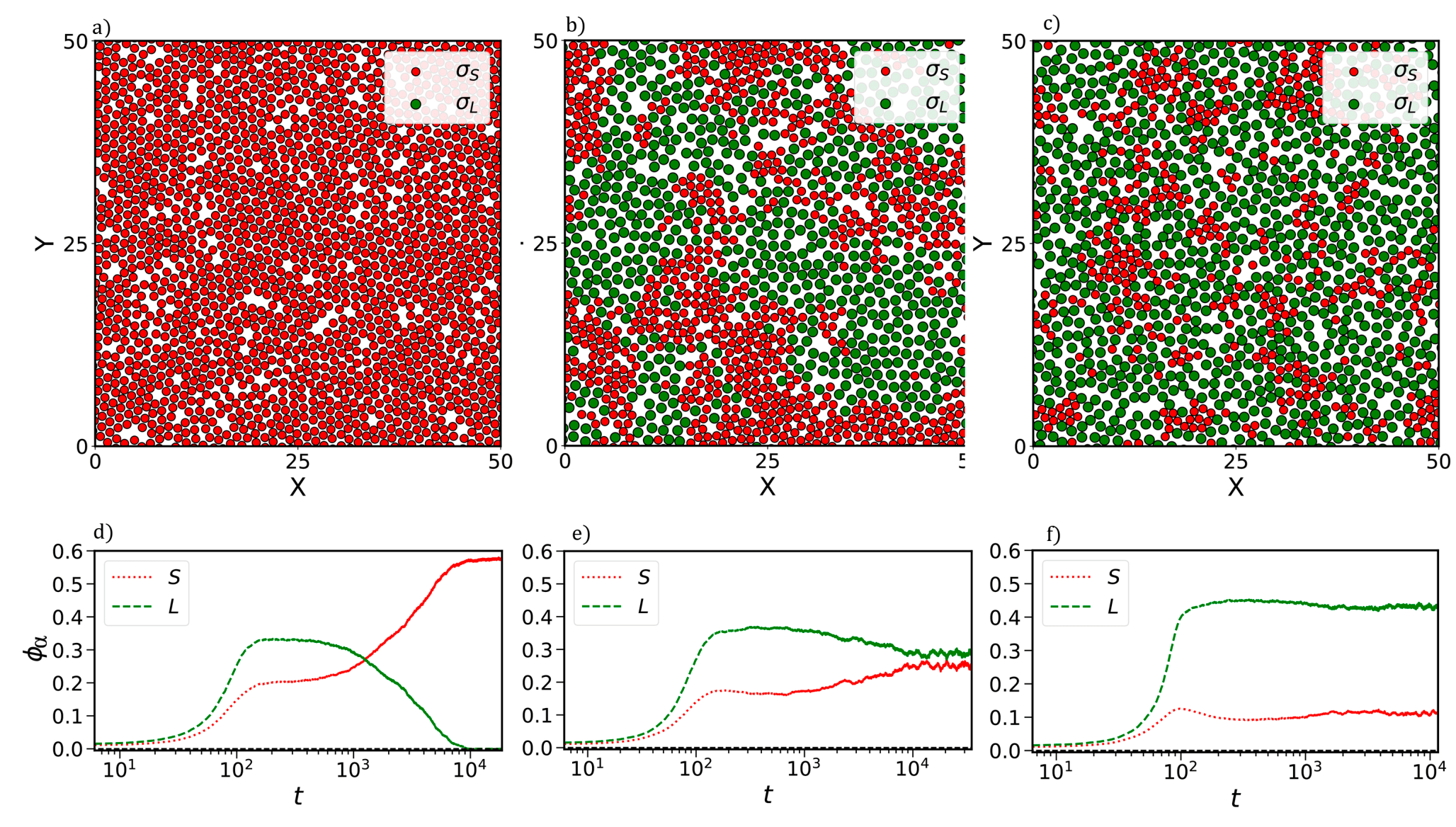} 

  \caption{Panels in the first row [a)-d)] show snapshots 
  of the spatial distribution of passive disks space at long times
  (graphs only show a part of the full system, which has lateral size $L_s=150$). 
  Panels in the second row [e)-h)] show the corresponding time evolution of the packing fractions for both species,
  starting with a random configuration of $250$ disks of each type. 
  In the different columns we have used different values of $D_L$ and $D_S=0.001$: $D_L$ = 0.01, 0.02, 1.0 from left to right. Remaining 
parameters are $\delta=0.01$, $\beta=0.05$, $\sigma_L/\sigma_S=1.2$.}
  \label{Fig:Plots_Dif}
\end{figure*}

In this subsection we study
the case in which both
types of passive disks
 have the same birth and death rates but
 where
 their diffusion coefficients are different. 
 In biological applications
situations of this type have given
rise to contradictory results (albeit under
very different settings) so that 
 diffusing faster provides a competitive
advantage 
\cite{Pigolotti2014}
or can be detrimental \cite{Heinsalu2012}). 
Thus, one of the main 
questions we address is if high 
or low diffusivity can provide
 a selective advantage 
 for either of the two competing
 populations.
 In our model diffusivity
introduces a spatial scale increasing the 
effective size of the disks, and hence affecting 
the dynamics of space-limited growth.  

\begin{figure}
    \includegraphics[width=0.99\linewidth]{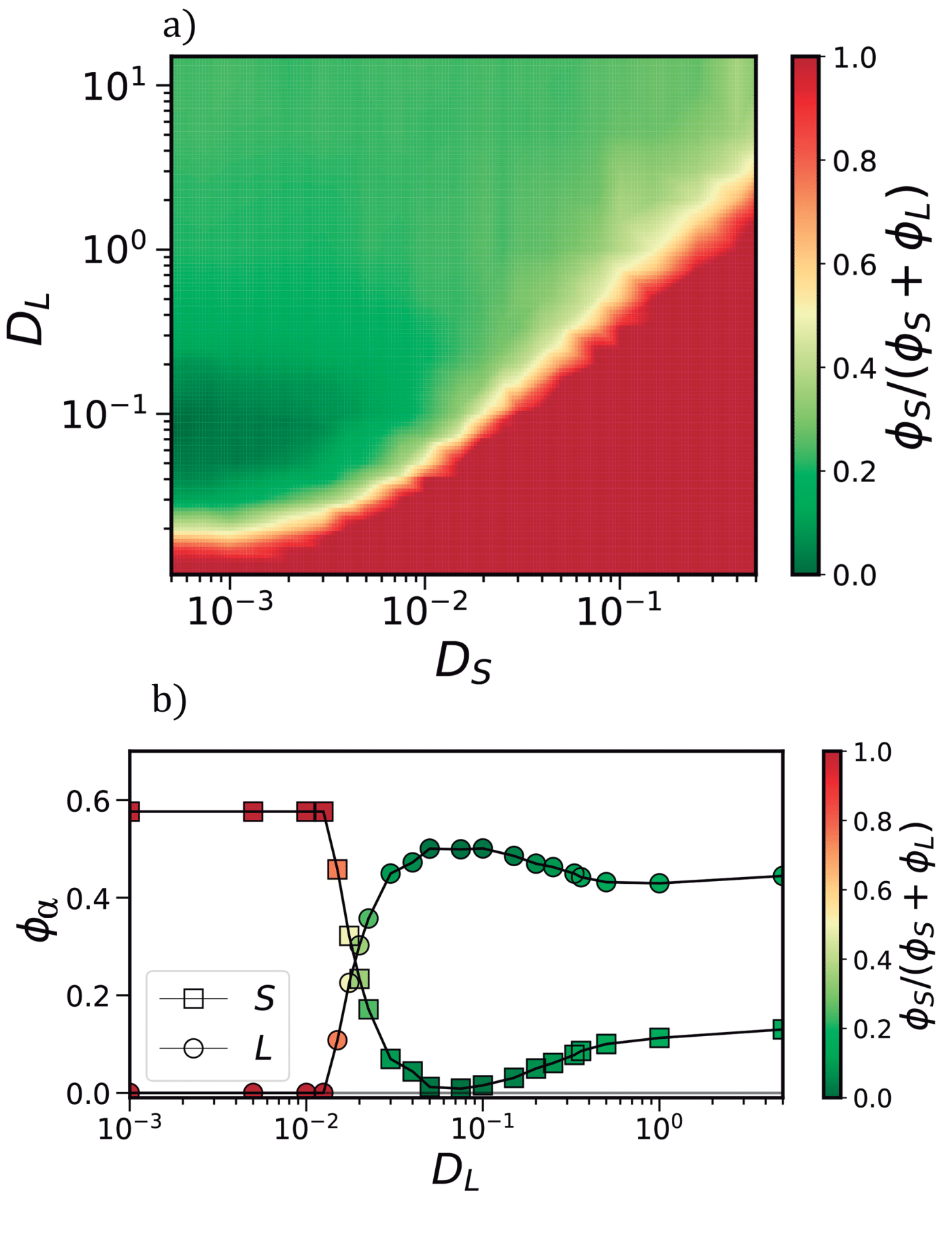}

  \caption{Phase diagram in the space 
  of diffusivities, $D_S$ and $D_L$.
   Panel a): Background color 
   indicates the relative filling fraction 
   of the two type of disks.
    In the red region the smaller 
   particles fill most of the space, 
   in the green region the 
   larger particles dominate.
  Panel b): Long-time average packing fractions 
  for both species as 
  a function of $D_L$, for 
  fixed $D_S=0.001$. The colors 
  of the symbols indicate
   the relative amount 
  of space occupied 
  by the small particles. 
  Model parameters are
  $\delta=0.01$, $\beta_L=\beta_S=0.05$, $\sigma_S=1.0, \sigma_L=1.2$.}
  \label{Fig:Diagram_Dif}
\end{figure}

In the upper row  
of  Fig.~\ref{Fig:Plots_Dif}
 we show
the spatial distribution of disks at 
long times,
for different $D_L$ and leaving all
other model parameters fixed.
In the leftmost panel (small ratio $D_L/D_S$) 
the smaller particles dominate the system. 
In the central panel (intermediate $D_L/D_S$),  
there is a balance of both types of disks, 
and  when $D_L/D_S$ is sufficiently high 
(rightmost panel), the larger 
disks occupy most of the space.
 Figs.~\ref{Fig:Plots_Dif} d)-f) 
  show the time evolution of the packing fractions 
  of both types in simulations
  starting with $250$ particles of each size.
The corresponding phase diagram, 
for  $\sigma_L/\sigma_S=1.2$, 
is shown in
 Fig. \ref{Fig:Diagram_Dif}a). 
Increasing the diffusivity 
of either species promotes
an increased relative 
abundance of that species. 
This is because larger diffusivity
for a given particle means
it effectively occupies more space
where  others cannot 
place their offspring.
In particular we observe a 
transition between 
phases in which the system is 
predominantly filled by 
the smaller particles 
to one in which it is 
mostly filled by larger particles.

\begin{figure}
    \includegraphics[width=0.95\linewidth]{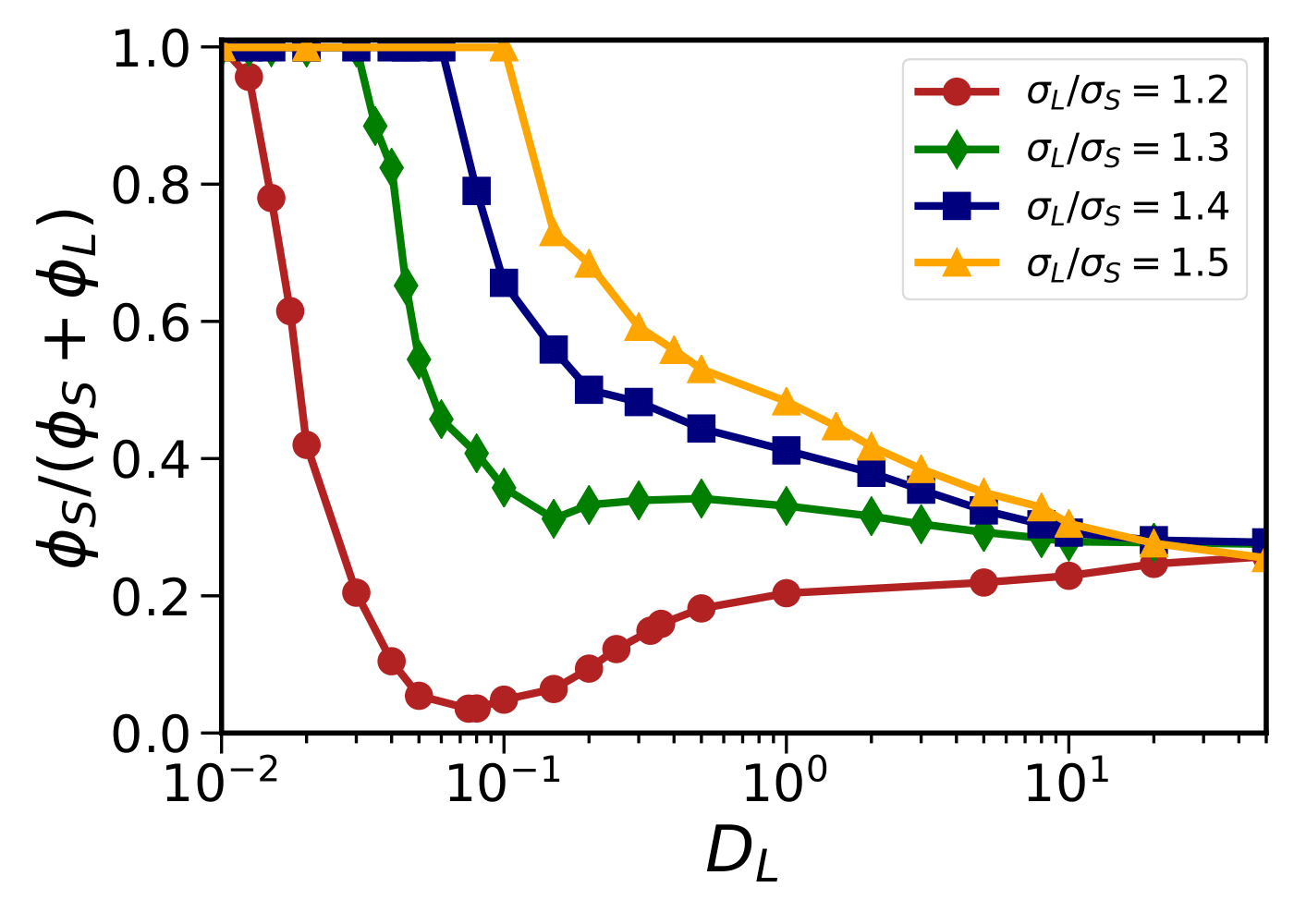}
  \caption{Normalized packing fraction of
   small particles versus $D_L$ for
  different values of $\sigma_L/\sigma_S$ as indicated. 
  Same remaining parameters as in Fig. \ref{Fig:Diagram_Dif}}.
  \label{Fig:Size_Dif}
\end{figure}

We conclude that increasing diffusivity 
of the larger disks can reverse the 
competitive advantage of the smaller 
ones. At difference with the results
of the previous subsection (Fig.~\ref{Fig:Plots_Birth}), 
the dominance of  larger disks 
for high $D_L/D_S$ is not complete though. 
Instead, we  find a remaining population 
of smaller disks occupying about $10 \mhyphen 15\%$ of
 space until the end of  our simulations
 (i.e., $\phi_S=0.1 \mhyphen 0.15$),
see Fig.~\ref{Fig:Diagram_Dif}b). We attribute 
this to interstitial holes, i.e., empty space 
between the disks, 
noting that the effect is more pronounced 
(i.e., $\phi_S/(\phi_S+\phi_L)$ becomes larger)
when the ratio $\sigma_L/\sigma_S$ is increased
 (see Fig.~\ref{Fig:Size_Dif}). 


An exception to the survival
of the smaller disks
can be observed in 
a region in $(D_S, D_L)$ space 
 in which the larger disks 
almost completely 
 dominate the system [dark green
 to the lower left in
Fig.~\ref{Fig:Diagram_Dif}a)]. 
This dominance
 only occurs when $D_S$
is small enough so that the movement  of the 
smaller disks is negligible 
with respect to their lifetime.
This region  disappears as the ratio
$\sigma_L/\sigma_S$ becomes larger 
(see Fig. \ref{Fig:Size_Dif}),
so that we think it is because 
small disks may occupy
the 
interstitial holes.





\subsection{Active particles}
\label{sub:active}

We consider systems 
in which both types 
of disks are self-propelled,
possibly with different 
values of the propulsion
velocities
 $v_L$ and $v_S$. 
The remaining 
parameters are set to
the same: $\beta_L=\beta_S=\beta$, $D_L=D_S=D$, and
$\delta=0.01$.

We first take a 
birth rate of $\beta=0.05$, 
and diffusivity of  $D=0.05$. 
Under these conditions, without activity, the system 
  is in a liquid state, which means that 
  the packing fraction of both types 
  is low and there is no ordering. 
 Also the system reaches
the steady state faster.
Our motivation for this choice is that
 we 
aim at analysing MIPS, which is typically not observed
   for large values  of the packing fraction (solid phase)
\cite{Almodovar2022,digregorio2018}.

In the previous section
we have shown that 
Brownian mobility provides 
an advantage in the
 competition for space 
between the two types of disks.
Given that self-propulsion yields
comparable effects (at low velocities)
to 
 diffusivity \cite{Levis2015,Palacci2010}
 we expect in this low activity regime
similar outcomes for active
particles. 
However, MIPS typically 
appears when activity is large,
and thus we study this regime
in detail by varying 
the self-propulsion 
coefficients of both types of disks,
and compute the average packing 
fraction of each species at long times.

\begin{figure}
    \includegraphics[width=0.99\linewidth]{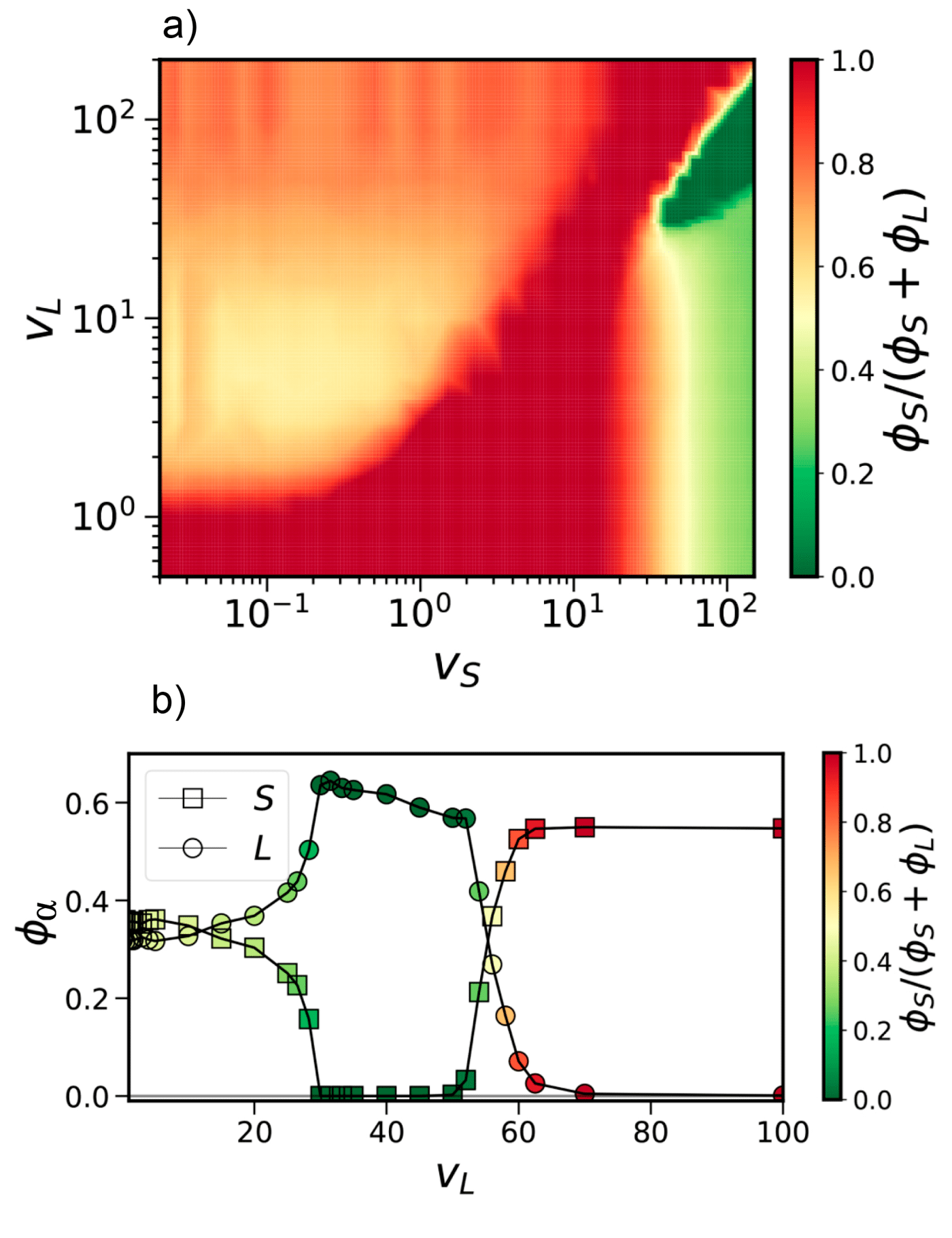}
  \caption{Panel (a): Phase diagram
   of the system with active disks in the $(v_L, v_S)$ plane. Background color
  represents the normalised packing fraction of the
smaller particles (darker red color indicates dominance
 of the smaller type, darker green 
  dominance of the larger).
Panel (b): Long-time average packing fractions for both species as function
of $v_{L}$ for fixed $v_{S}=50$. The colors of the symbols indicate
 the relative  space occupied by the small particles. 
 Model parameters are $\delta=0.01$, $\beta_L=\beta_S=0.05$, $\sigma_S=1.0, \sigma_L=1.2$.}
\label{Fig:Diagram_Act}
\end{figure}

The resulting phase diagram, obtained
for $\sigma_L/\sigma_S=1.2$,
 is shown in
Fig.~\ref{Fig:Diagram_Act}a). When the activity
of both types of disks is
sufficiently 
 low 
 the smaller disks dominate the system. As
$v_L$ increases (keeping $v_S$ fixed at a
 sufficiently small value) a transition 
 to coexistence is found. This coexistence 
remains as the activity of 
the larger particles is further increased. 
We find  interesting behaviour
for sufficiently large  $v_S$. 
This is
the region of 
MIPS.  
We observe that  coexistence can be 
reached for low and intermediate 
values of $v_L$ (see Fig. Fig.~\ref{Fig:Diagram_Act} b)). 
In between there is a range of values
where large particles dominate, and for very high
$v_L$ small particles dominate the system.

\begin{figure*}[hbt!]
  \centering
    \includegraphics[width=.99\linewidth]{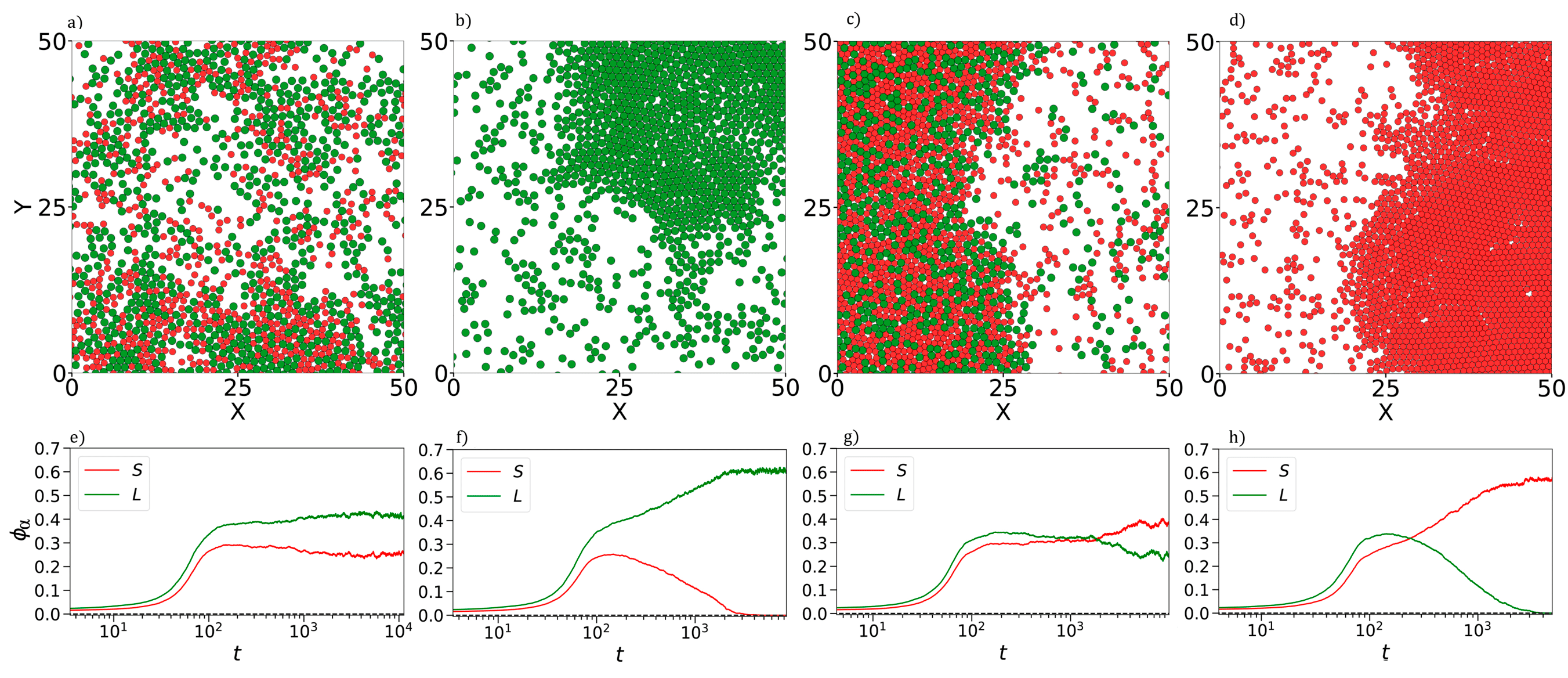} 
  \caption{Panels in the first row (a-d) show snapshots 
  of the spatial distribution of active disks space at long times
  (only a part of the full size $L_s=150$). 
  Panels in the second row (e-h) correspond 
  to the temporal evolution of the packing fraction
  starting with a random configuration of $250+250$ disks. 
  In addition, for each column, we have used different values of $v_L$ and $v_S=50$. 
  The values used are $v_L$ = 25, 40, 56 and 70 from left to right.}
  \label{Fig:Plots_Act}
\end{figure*}

Which type of particle survives
(or the presence of coexistence) results
from a complicated interplay between
the separation in dense and diluted 
phases
formed of
both species in non-steady conditions
because of activity, and
the birth/death dynamics which is itself mediated
by the spatial distribution of disks.   
This is shown in the upper panels of 
Fig.~\ref{Fig:Plots_Act} 
with some examples 
of the disk distribution  
at long times for $v_S=50$, 
and different values of $v_L$. 
In  panel a)  there is coexistence
 for sufficiently small $v_L$.  
As we increase $v_L$ the 
smaller species becomes extinct 
(dark green region in the phase diagram) 
[Fig.~\ref{Fig:Diagram_Act}b)].
As $v_L$ is increased further,
 coexistence is observed 
again and both species show MIPS 
[see Fig. \ref{Fig:Plots_Act}c)]. 
 For even higher values of $v_L$, the  large disks
 become extinct
  [see Fig.~\ref{Fig:Plots_Act}d)]. In the lower row we show 
  the corresponding time evolutions 
 of the packing fractions of both types of particles
 starting from a configuration with $250$ particles of
 each.


\begin{figure}
\includegraphics[width=0.95\linewidth]{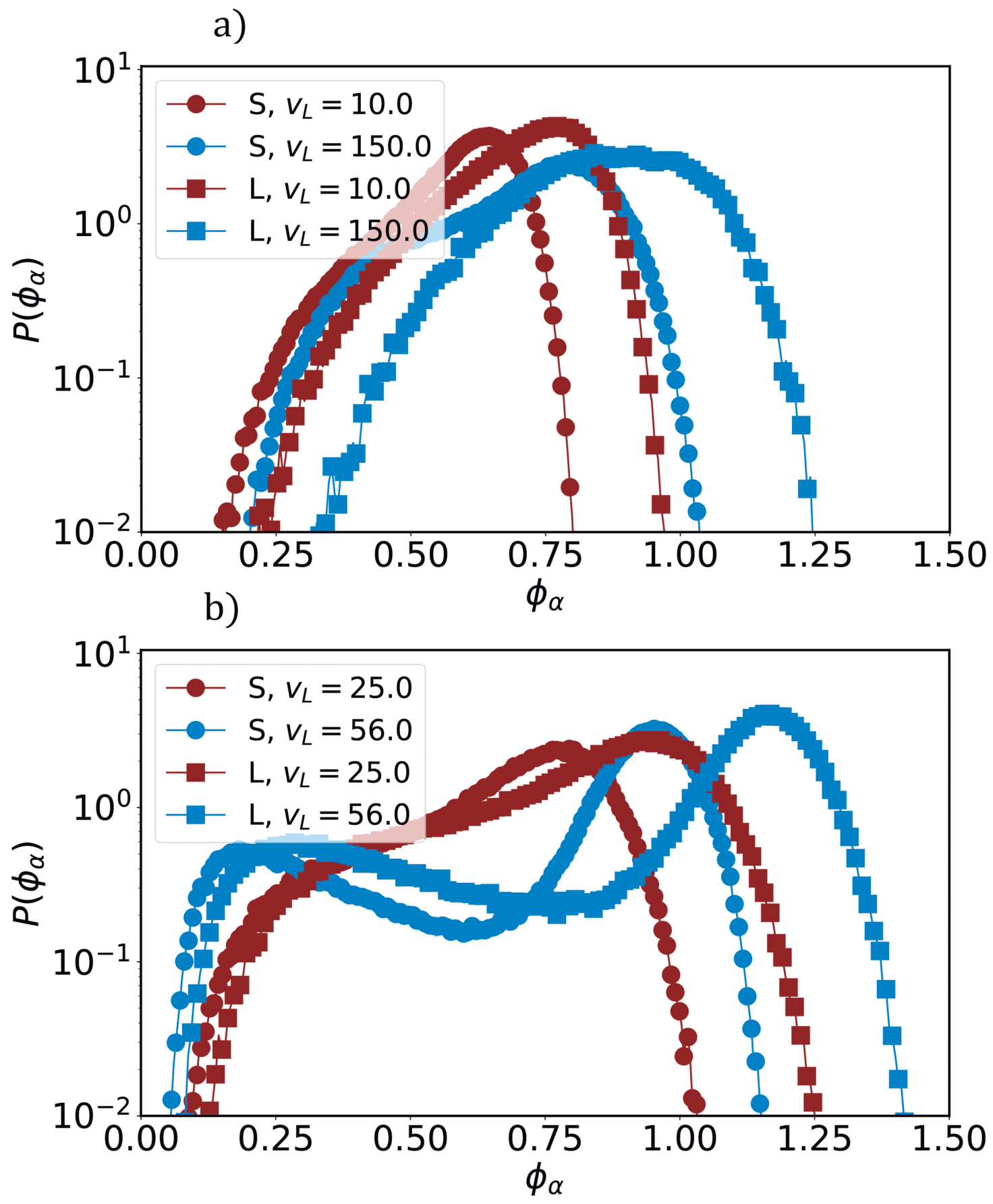}
\caption{Distribution of the local packing fraction for different values of activity
in configurations where both disks types coexist.
 Panel (a): $v_S=0.1$. Panel (b): $v_S=50$. 
 The  $v_L$ value is shown in the legend. 
   }
\label{Fig:MIPS}
\end{figure}

Let us next characterise the emerging 
spatial structures 
when there is coexistence. 
To do this we compute the 
distribution of local 
packing fractions
 of large disks,
$P(\phi_L)$, and
small ones, $P(\phi_S)$.
These are obtained from
the local packing fraction
of either species at each location in 
space.   
MIPS is characterised by a double-peaked 
distribution,
 with the two 
peaks corresponding 
to the dense and diluted 
phases, respectively \cite{digregorio2018}.
In Fig.~\ref{Fig:MIPS}a) we plot
 $P(\phi_S)$
and $P(\phi_L)$ corresponding to 
two situations: i) small particles with 
very low $v_S=0.1$, and large particles
with low $v_L=10$ (the two distributions
are plotted in red); ii) small particles
again with  $v_S=0.1$, but large particles
with high $v_L=150$. These scenarios correspond,
respectively, to the cases
where (if the two types were in isolation without
the presence of the other) 
a) neither small nor large disks would form
MIPS, and b) small disks would not form MIPS,
but large ones would because they
have a high $v_L$. 
However 
when both types of particles are present in the system,
the distribution of local packing
fractions is single-peaked, and
there is no separation
between dilute
and dense phases.

Instead, in  Fig.~\ref{Fig:MIPS}b) we 
use a higher value of $v_S=50$, and two different
values of $v_L$:
$v_L=25$ for which there would be no MIPS for
the large particles in isolation,
and $v_L=56$, for which there would be MIPS.
Now  $P(\phi_{L})$
and $P(\phi_{S})$ are double-peaked, that is
both types of disks have a dense and a dilute phase.
These results indicate that
the two species arrange in the same 
 spatial distribution when
they coexist.
Either both display a single phase, or both
species show MIPS.

\begin{figure}
\includegraphics[width=0.98\linewidth]{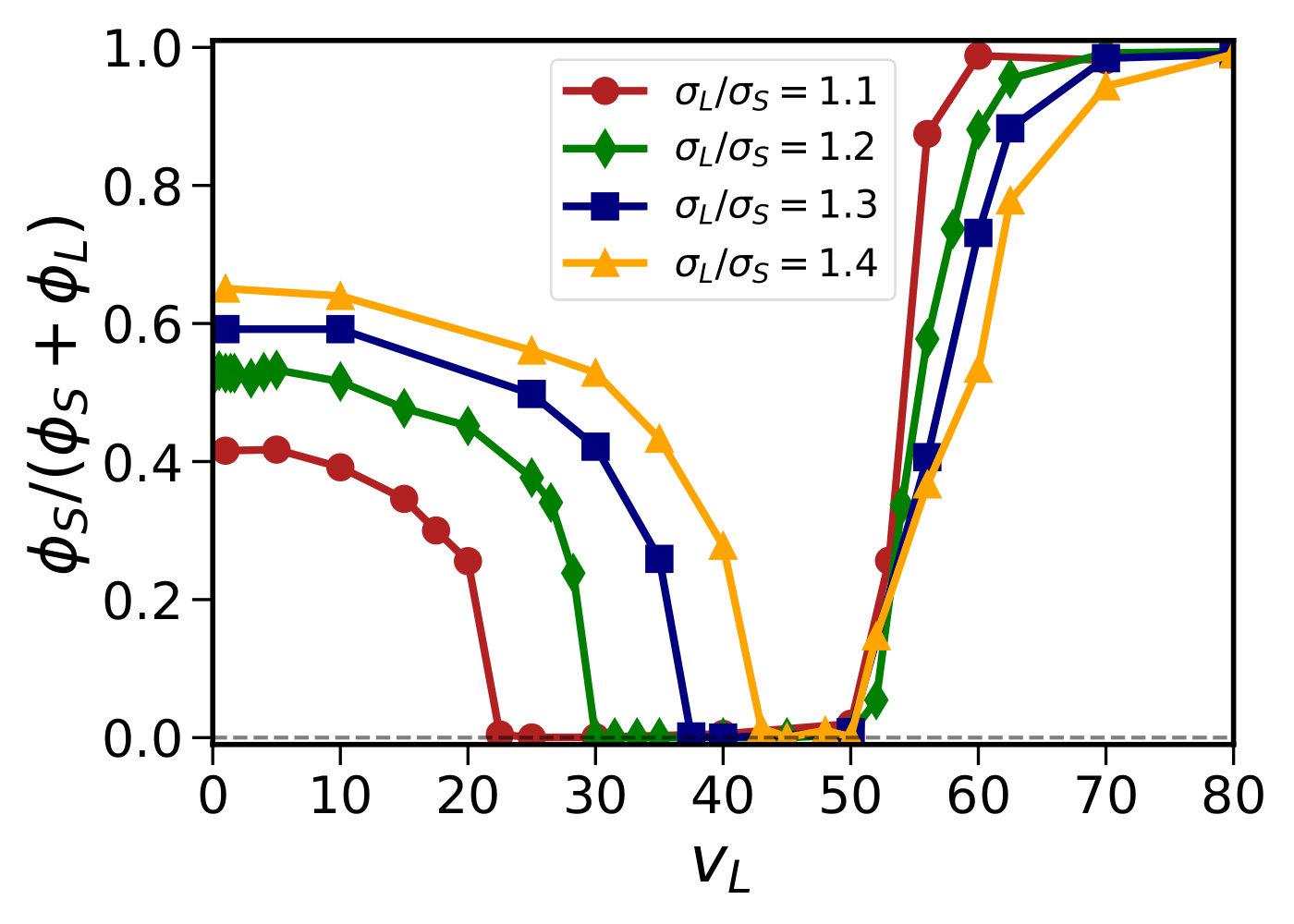}
\caption{Normalized packing fraction of the smaller type of particle versus $v_L$
 for different values of $\sigma_L/\sigma_S$ as indicated in the legends.
 $v_S= 50$, same remaining  parameters as in Fig. \ref{Fig:Diagram_Act}.}
\label{Fig:Size_Act}
\end{figure}

Finally, in Fig. \ref{Fig:Size_Act}
we plot the normalized packing fraction 
for the smaller types of disks
when $v_L$ varies (keeping 
$v_S=50$ fixed, i.e., within the MIPS region)
and 
 for several values of the ratio
 of size particles.  
When we increase $\sigma_L/\sigma_S$, the effect
 of the interstices becomes more pronounced
 which 
  causes the small particles 
 packing fraction to increase.
 For the same reason the 
 region in  Fig. \ref{Fig:Size_Act} 
in which the smaller particles 
go extinct reduces. 
In other words, as $\sigma_L/\sigma_S$ 
increases, the large particles need a higher
value of activity to form MIPS. This 
results in a higher value of $v_L$
for the small particles to go extinct.
%

\section{Summary and discussions}
\label{Sec:summary}

The interplay between
motility and proliferation
is fundamental for many biological
processes.
In this work we have studied
a binary mixture of two
types of 
motile disks with
different sizes, and
undergoing birth 
and death dynamics.
At difference with our previous 
work \cite{Almodovar2022}
where we concentrated on 
the spatial structure
formed by a single
population of identical disks,
in this work we have 
studied the conditions
for coexistence and 
which of the two types
does not go extinct in the long-time.
Since the birth probability is limited by
size, smaller disks
have larger chances to
reproduce (when the remaining characteristics
are identical). Thus we focused
on the conditions
under which this is reversed, i.e.,
larger disks survive,
or there is
coexistence of both types
of disks.
We have considered two main cases:
passive  and self-propelled disks.
For passive particles we 
have discussed 
the
role of growth rates and diffusivities,
and have obtained typical phase diagrams
when the ratio of particles sizes
is $\sigma_L/\sigma_s=1.2$, but discussed also
it for other ratios, unveiling  the role of
interstices.
We observe that Brownian mobility provides
with an advantage such that the larger
particles, with less chances to reproduce,
can coexist with the smaller or even dominate
in the steady state when they diffuse faster. 

We have analysed the role
of activity and 
have shown that,
as expected, it is
similar similar to that of diffusivity when self-propelled velocities
have low values. However, when MIPS is present
(typically obtained for larger values of activities)
the coexistence dynamics changes, and the typical
situation is a coexistance 
of both species, and both showing
the same diluted and dense phases. 

Further work should be 
devoted to analyse 
the spatial structure 
 in the different situations we have described,
and in particular, the 
influence on the hexatic and solid
phases, if any, of 
the binary mixture with demography.
Concerning biological applications, the main
motivation for our work, 
where populations
are characterized by large diversity,
it is of great interest to
consider not only two disks sizes but
a whole size distribution, 
but also of the other
parameters like the growth 
rates and the self-propulsion
velocities.

\acknowledgments

A.A and C.L. acknowledge grant
LAMARCA PID2021-123352OB-C32 
funded by MCIN/AEI/10.13039/501100011033323
and FEDER “Una manera de hacer Europa”.
T.G. acknowleges
 partial financial support from the Agencia Estatal
  de Investigaci\'on and Fondo Europeo de Desarrollo 
  Regional (FEDER, UE) under project APASOS 
  (PID2021-122256NB-C21, PID2021-122256NB-C22), 
  and the Maria de Maeztu programme for Units of Excellence, 
  CEX2021-001164-M funded by  MCIN/AEI/10.13039/501100011033.

\appendix

\section{}
\label{ApA}

\begin{figure*}[hbt!]
  \centering
    \includegraphics[width=.99\linewidth]{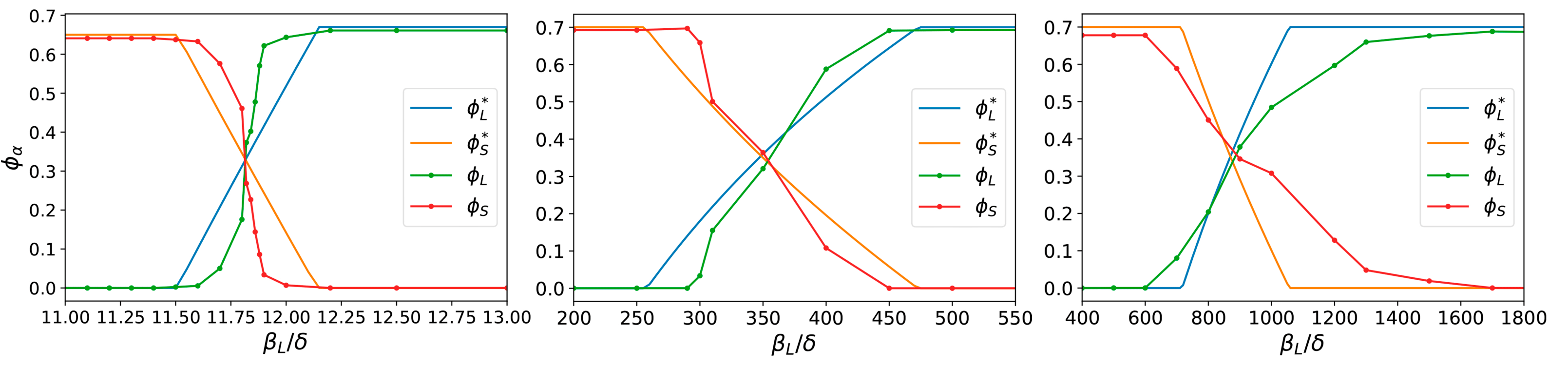} 
  \caption{Long-time average packing fractions for both species as a function of $\beta_L/\delta$, for fixed $\beta_S=0.1$ (left) and $\beta_S=0.85$ (middle and right), and $\delta=0.01$. $\phi_{\alpha}^{*}$ obtained from Eq.~(\ref{pfnullclines}) and $\phi_{\alpha}$ from particle simulations. From left to right: $\sigma_L/\sigma_S=1.2, 1.3, 1.4$. }
  \label{Fig:Plots_ActApp}
\end{figure*}


The parameters for which both species 
coexist can be approximated 
from the non-zero fixed points of
 Eqs.~(\ref{eq:competitionpacking}):
\begin{eqnarray}
 \phi^*_S = \kappa_S - \alpha_S \phi^*_L \left(\frac{\sigma_S}{\sigma_L}\right)^2, \nonumber \\
 \phi^*_S = \frac{1}{\alpha_L}( \kappa_L- \phi^*_L)\left(\frac{\sigma_S}{\sigma_L}\right)^2,
 \label{eq:nulclines}
\end{eqnarray}
with 

\begin{eqnarray}
\kappa_S &=& \frac{a \beta_S-\delta}{a\beta_S} \phi_{S,{\rm max}}, \nonumber\\
    \kappa_L &=& \frac{a \beta_L-\delta}{a\beta_L} \phi_{L,{\rm max}}. 
\end{eqnarray}

Both equalities are fulfilled at the same time in the coexistence region. Furthermore, recall that $\phi_{S,\rm max}=\phi_{L,\rm max}$, since the maximum value does not depend on either the type of particle or the size of the system. Now,  we make the ansatz:

\begin{eqnarray}
\alpha_S=c_S(\beta_L/\beta_S)^\gamma,\nonumber\\
\alpha_L=c_L(\beta_S/\beta_L)^\gamma,
\label{Ap:ansatz}
\end{eqnarray}
where $c_S$, $c_L$ and $\gamma$ are constants. 

We have assumed  symmetry in the interaction
of the species, so that both species have
the same exponent $\gamma$. 
Inserting the ansatz (\ref{Ap:ansatz}) in  
Eq.~(\ref{eq:nulclines}) we obtain 
an expression for the packing fraction with explicit dependence
 on the demographic rates:

\begin{eqnarray}
\phi^*_S=\frac{1}{1-c_L c_S}\left(\kappa_S-\kappa_L c_S\left(\frac{\beta_L}{\beta_S}\right)^\gamma \left(\frac{\sigma_S}{\sigma_L}\right)^2\right), \nonumber\\
\phi^*_L=\frac{1}{1-c_L c_S}\left(\kappa_L-\kappa_S c_L\left(\frac{\beta_S}{\beta_L}\right)^\gamma \left(\frac{\sigma_L}{\sigma_S}\right)^2\right).
\label{pfnullclines}
\end{eqnarray}

These expressions predict 
 the 
value of the steady packing fractions
 at each point of the diagram $(\beta_L,\beta_S)$ inside the 
 coexistence region.  In the  region 
 where only large (small) 
 particles are found, we have 
 $\phi^*_L=\kappa_L$ and $\phi^*_S=0$
  ($\phi^*_L=0$ and $\phi^*_S=\kappa_S$).

We have carried out a series of numerical 
experiments for various values of $\delta$, $\sigma_L$, $\sigma_S$,  (not shown) 
beyond those used in
 Fig. \ref{Fig:Diagram_Birth}
and have observed that 
the phase diagram is almost
 independent of the value of $\delta$ (always 
 assuming $\beta>\delta$), 
 and the ratio $\sigma_L/\sigma_S$.
The coefficients $c_L$, $c_S$ and $\gamma$ in Eq. (\ref{pfnullclines})
can be obtained from fits of the packing fraction
(see Fig.~\ref{Fig:Plots_ActApp}). 
We compute the difference between 
 the packing fraction obtained from Eq. (\ref{pfnullclines}),
and the one obtained from the numerical simulations 
of the particle model 
for different values of 
$\delta$, $\sigma_S$, $\sigma_L$, $\beta_S$, $\beta_L$. Then, 
we minimize the sum of all the differences.

 From these simulations, 
the best fit is obtained
 for $c_L=0.69$, $c_S=1.34$ and $\gamma= 1/8$.

\newpage
\nocite{}
\bibliography{bibl.bib}

\end{document}